\documentclass[%
 reprint,
 amsmath,amssymb,
 aps,
]{revtex4-1}

\usepackage{graphicx}
\usepackage{dcolumn}
\usepackage{bm}
\usepackage{multirow}
\usepackage{blindtext}

\begin{document}

\preprint{APS/123-QED}

\title{Moments of the neutron $g_2$ structure function at intermediate $Q^2$}

\author{P.~Solvignon$^{1}$, N.~Liyanage$^{2}$, J.-P.~Chen$^{3}$, Seonho~Choi$^{4}$, 
K.~Slifer$^{1}$, K.~Aniol$^5$, T.~Averett$^{6}$, W.~Boeglin$^7$, A.~Camsonne$^3$, 
G.~D.~Cates$^{2}$, C.~C.~Chang$^{8}$, E.~Chudakov$^{3}$, B.~Craver$^{2}$, F.~Cusanno$^9$, 
A.~Deur$^{3}$, D.~Dutta$^{10}$, R.~Ent$^{3}$, R.~Feuerbach$^{3}$, S.~Frullani$^9$, H.~Gao$^{10}$, 
F.~Garibaldi$^9$, R.~Gilman$^{11}$, C.~Glashausser$^{11}$, V.~Gorbenko$^{12}$, O.~Hansen$^3$, 
D.~W.~Higinbotham$^3$, H.~Ibrahim$^{13}$, X.~Jiang$^{11}$, M.~Jones$^{3}$, A.~Kelleher$^{6}$, 
J.~Kelly$^{8}$\footnote[2]{Deceased}, C.~Keppel$^{3}$, W.~Kim$^{14}$, W.~Korsch$^{15}$, K.~Kramer$^{6}$, 
G.~Kumbartzki$^{11}$, J.~J.~LeRose$^{3}$, R.~Lindgren$^{2}$, B.~Ma$^{16}$, D.~J.~Margaziotis$^5$, 
P.~Markowitz$^7$, K.~McCormick$^{11}$, Z.-E.~Meziani$^{17}$, R.~Michaels$^{3}$, 
B.~Moffit$^{6}$, P.~Monaghan$^{16}$, C.~Munoz Camacho$^{18}$, K.~Paschke$^2$, B.~Reitz$^3$, 
A.~Saha$^3$\footnotemark[2], R.~Shneor$^{19}$, J.~Singh$^{2}$, V.~Sulkosky$^{6}$, 
A.~Tobias$^{2}$, G.~M.~Urciuoli$^{20}$, K.~Wang$^{2}$, K.~Wijesooriya$^{10}$, 
B.~Wojtsekhowski$^{3}$, S.~Woo$^{14}$, J.-C.~Yang$^{21}$, X.~Zheng$^2$, L.~Zhu$^{16}$\\
~}

\collaboration{Jefferson Lab E01-012 Collaboration}

\affiliation{$^1$University of New Hampshire, Durham, NH 03824}
\affiliation{$^2$University of Virginia, Charlottesville, VA 22904}
\affiliation{$^3$Thomas Jefferson National Accelerator Facility, Newport News, VA 23606}
\affiliation{$^4$Seoul National University, Seoul, 151-747, Korea}
\affiliation{$^5$California State University, Los Angeles, Los Angeles, CA 90032}
\affiliation{$^6$College of William and Mary, Williamsburg, VA 23187}
\affiliation{$^7$Florida International University, Miami, FL 33199}
\affiliation{$^8$University of Maryland, College Park, MD 20742}
\affiliation{$^9$Istituto Nazionale di Fisica Nucleare, Gruppo Collegato Sanit\'{a}, Seziona di Roma, 00161 Roma, Italy}
\affiliation{$^{10}$Duke University, Durham, NC 27708}
\affiliation{$^{11}$Rutgers, The State University of New Jersey, Piscataway, NJ 08855}
\affiliation{$^{12}$Kharkov Institute of Physics and Technology, Kharkov 61108, Ukraine}
\affiliation{$^{13}$Cairo University, Giza 12613, Egypt}
\affiliation{$^{14}$Kyungpook National University, Taegu City,  Korea}
\affiliation{$^{15}$University of Kentucky, Lexington, KY 40506}
\affiliation{$^{16}$Massachusetts Institute of Technology, Cambridge, MA 02139}
\affiliation{$^{17}$Temple University, Philadelphia, PA 19122}
\affiliation{$^{18}$Universit\'{e} Blaise Pascal et CNRS/IN2P3 LPC, 63177 Aubi\`{e}re Cedex, France}
\affiliation{$^{19}$University of Tel Aviv, Tel Aviv, 69978 Israel}
\affiliation{$^{20}$Istituto Nazionale di Fisica Nucleare, Sezione di Roma, 00185 Roma, Italy}
\affiliation{$^{21}$Chungnam National University, Taejon 305-764, Korea}

\date{\today}

\begin{abstract}
We present new experimental results for the $^3$He spin structure function $g_2$ in the resonance region at $Q^2$ values 
between 1.2 and 3.0 (GeV/c)$^2$. Spin dependent moments of the neutron were  extracted. Our main result, the inelastic 
contribution to the neutron $d_2$ matrix element, was found to be small at $\langle Q^2 \rangle = 2.4~\rm{(GeV/c)}^2$ and in 
agreement with the Lattice QCD calculation. The Burkhardt-Cottingham sum rule for $^3$He and the neutron was tested with 
the measured data and using the Wandzura-Wilczek relation for the low $x$ unmeasured region. 

\begin{description}
\item[PACS numbers]
13.60.Hb, 13.88.+e, 14.20.Dh
\end{description}
\end{abstract}

\pacs{Valid PACS appear here}
\maketitle


\section{ Introduction} 
The internal structure of a nucleon probed in inclusive scattering  can be expressed in terms of four structure functions: two unpolarized structure 
functions ($F_1$ and $F_2$) and two polarized structure functions ($g_1$ and $g_2$). Within the Quark-Parton Model 
$F_1$, $F_2$ and $g_1$ depend on unpolarized and polarized quark distributions. In contrast, $g_2$ has no direct link 
to quark distributions but is related to the interaction between quarks and gluons inside the nucleon. This makes the 
$g_2$ structure function ideal for the study of quark-gluon correlations. 

The measurements of nucleon polarized structure functions in Deep Inelastic Scattering (DIS) have been instrumental in 
advancing our understanding of Quantum Chromo-Dynamics (QCD) (for a recent review of nucleon spin structure 
measurements, see~\cite{Kuhn:2008sy,Chen:2010qc}). In DIS, the incident electron interacts with the nucleon constituents 
by exchanging a virtual photon of four-momentum squared $q^2= - Q^2$ and energy $\nu$. At very large values of 
$Q^2$, the lepton-nucleon interaction can be described by the incoherent sum of quasi-elastic scattering from 
asymptotically free quarks, with a momentum fraction  $x=Q^2/(2M\nu)$ of the parent nucleon's momentum ($M$ is the 
mass of the nucleon). Most of the former polarized structure function measurements  were performed using nucleon targets polarized longitudinally with 
respect to the lepton spin. In this case the helicity dependent cross section difference is dominated by the $g_1$ spin 
structure function, and as a result, this structure function is known with high precision in most kinematic regions. 

In the  Quark-Parton Model, the contributions  to the structure functions due to   electron scattering off  the asymptotically free quarks inside the nucleon  are independent of $Q^2$, up to corrections due to gluon radiation and  vacuum polarization. At high $Q^2$ these corrections can be accurately calculated using perturbative QCD. As  $Q^2$ decreases, quark-gluon and quark-quark correlations make 
increasingly important contributions to the structure functions. In the $g_1$ structure function these correlation terms are 
suppressed by factors of $(1/Q)^n$ with respect to the asymptotically free contributions. In the case of the second spin 
structure function, $g_2$, the non-perturbative multi-parton correlation effects contribute at the same order in $Q^2$ as 
asymptotically free effects. 

The moments of structure functions provide especially powerful tools to study fundamental properties of the nucleon because they can be compared to rigorous theoretical results like sum rules and Lattice QCD calculations. The Operator 
Product Expansion (OPE) of QCD~\cite{Wilson:1969zs,Kodaira:1979ib} can be used to relate the hadronic matrix elements 
of current operators to the  moments of structure functions.  In the OPE, the moments are expanded  in a series ordered 
by $1/Q^{\tau -2}$. In this expansion $ \tau = 2,3,4....$  is known as the twist (dimension - spin)  of the operator. The 
twist-2 contributions to the moments correspond to scattering off asymptotically free quarks, where the higher twist 
contributions arise due to multi-parton correlations. 

The Cornwall-Norton (CN) moments of  $g_1$ and $g_2$ are defined by the equation:
\begin{eqnarray}
\Gamma^{(n)}_{1,2}(Q^2) \equiv \int_0^1 dx~x^{(n-1)}~g_{1,2}(x,Q^2). 
\label{eq:moments}
\end{eqnarray}
In addition, at high $Q^2$, the twist-3 reduced matrix element $d_2$ can be related to the second  moment of a certain 
combination of $g_1$ and $g_2$:
\begin{eqnarray}
d_2(Q^2) & = &  \int_0^1 dx~x^{2}~\left[2g_1(x,Q^2) + 3g_2(x,Q^2)\right] \nonumber \\
               & = &  3\int_0^1 dx~x^{2}~\left[g_2(x,Q^2) - g_2^{WW}(x,Q^2)\right].
\label{eq:d2}
\end{eqnarray}
Furthermore, the leading twist contributions to $g_2$ can be calculated using measured values of $g_1$ in the  
Wandzura-Wilczek relation, 
\begin{eqnarray}
g_2^{WW}(x,Q^2) = - g_1(x,Q^2) + \int_x^1 \frac{dy}{y} g_1(y,Q^2).
\label{eq:g2ww}
\end{eqnarray}
Hence, it is possible to cleanly isolate  the twist-3  contribution in a measurement of $g_2$ by subtracting the leading 
twist part.

\section{The experiment}
\begin{figure}[t]
\includegraphics[scale=0.65]{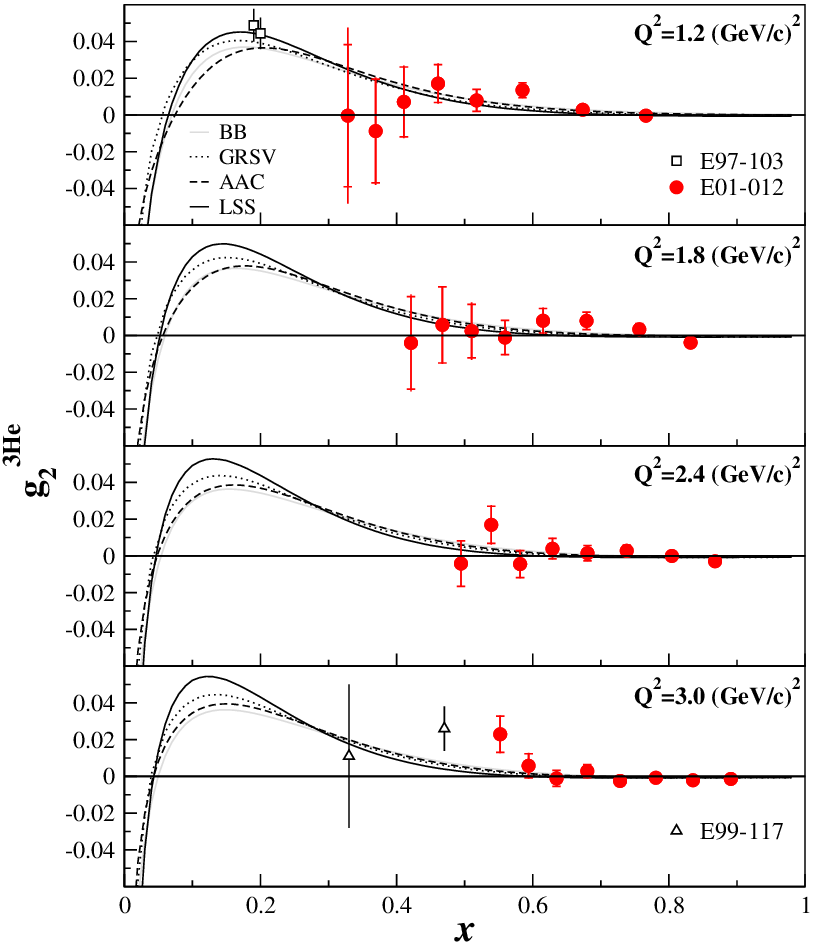}
\caption{\label{fig:g2} (Color online) The spin-structure function $g_2^{\rm ^3He}$ 
(per nucleon) in the resonance region at $Q^2$-values of 1.2, 1.8, 2.4 and 
3.0 (GeV/c)$^2$. The error bars represent the total uncertainties with the inner part being 
statistical only. Also plotted are the DIS JLab data from experiments 
E97-103~\cite{Kramer:2005qe} and E99-117~\cite{Zheng:2003un,Zheng:2004ce} (note 
that these data are at different $Q^2$). The curves were generated from the NLO parton 
distribution functions of Refs.~\cite{Blumlein:2002be,Gluck:2000dy,Hirai:2003pm,Leader:2005ci}.}
\end{figure}
The measurement of $g_2$ requires a longitudinally polarized electron beam scattering off a longitudinally and also 
transversely polarized nucleon according to the following formula:
\begin{eqnarray}
g_2 = \frac{M Q^2 \nu^2}{4 \alpha_e^2} \frac{1}{2 E^{\prime}} \frac{1}{E+E^{\prime}} 
\bigg[\frac{E+E^{\prime} \cos \theta}{E^{\prime} \sin \theta}\Delta \sigma_{\perp} - \Delta \sigma_{\parallel} \bigg]
\label{eq:g2}
\end{eqnarray}
where $\Delta \sigma_{\parallel}$ and $\Delta \sigma_{\perp}$ are the polarized cross section differences corresponding 
to longitudinal and transverse  target polarizations, respectively.  Their contributions to $g_2$ are weighted by three 
kinematical variables: the electron incident energy $E$, the scattered electron energy $E^{\prime}$ and angle $\theta$.  
The variable $\alpha_e$ is the electromagnetic constant. As can be seen in Eq.~\ref{eq:g2} the transverse polarized cross 
section difference is the dominant contribution to $g_2$. 
In the present paper we report results from Jefferson Lab (JLab)  Experiment E01-012 of the $g_2$ structure function measured in the nucleon resonance region 
at intermediate $Q^2$, using a polarized $^3$He target as an effective polarized neutron target.  We formed polarized 
cross-section differences from inclusive scattering of longitudinally polarized electrons off a longitudinally or transversely 
polarized $\rm{^3}$He target at a scattering angle of 25$^{\circ}$ for three incident beam energies, 3.028, 4.018 and 
5.009 GeV, and at 32$^{\circ}$ for an incident beam energy of 5.009 GeV. The beam polarization varied between 71.4 
and 84.9\% during the experiment depending of the incident beam energy and the running status of the other 
experimental halls.  The total relative uncertainty on the beam polarization was 3.4\%. The target spin could be set to 
three directions with respect to the beam helicity: two longitudinal configurations with target spin direction at 0 and 
180$^{\circ}$ and one transverse configuration at 90$^{\circ}$. The average target polarization was (38.0 $\pm$ 2.0)\% 
absolute. We used the two Hall A High Resolution Spectrometers in standard configuration for electrons 
detection~\cite{Alcorn:2004sb}. 
The structure function $g_2^{\rm ^3He}$ was extracted at constant beam energies and scattering angles. However, the 
integrations to form moments require the structure function values at a constant $Q^2$. Therefore our $g_2^{\rm ^3He}$ 
results were interpolated to extract $g_2^{\rm ^3He}$ values at four constant  $Q^2$-values  of 1.2, 1.8, 2.4 and 3.0 
$\rm{(GeV/c)}^2$. The results from $g_1^{\rm ^3He}$ were reported in a previous 
publication~\cite{Solvignon:2008hk,Solvignon:2006qc} along with the details of the experimental setup and the systematic 
uncertainties relevant to both structure functions. Figure~\ref{fig:g2} presents the results on $g_2^{\rm ^3He}$ from 
E01-012 at the four $Q^2$ values. Also shown are the $g_2^{\rm ^3He}$ curves generated from the NLO parton 
distribution functions of Refs.~\cite{Blumlein:2002be,Gluck:2000dy,Hirai:2003pm,Leader:2005ci} using 
Eq.~\ref{eq:g2ww}, including Target Mass Corrections (TMCs) from the formalism of Ref.~\cite{Sidorov:2006fi}.

\section{The twist-3 reduced matrix element $d_2$}
The  $g_2^{\rm ^3He}$ results at the four $Q^2$ values were used to evaluate the resonance region contribution to ${d}_2(Q^2)$ 
for $^3$He of Eq.~\ref{eq:d2}.  The DIS contribution at each  $Q^2$ value was evaluated from Eq.~\ref{eq:g2ww} with 
the already published E01-012 results of $g_1^{\rm ^3He}$~\cite{Solvignon:2008hk} . The $x$-region covered by our data 
corresponds to a range in the invariant mass of $1.080 \le W \le 1.905$ GeV at the given value of $Q^2$. 
Then $d_2(Q^2)$ for the  neutron was extracted from $d_2^{\rm ^3He}(Q^2)$ using the method described in Ref.~\cite{CiofidegliAtti:1996cg}:
\begin{eqnarray}
d_2^n = \frac{1}{p_n} d_2^{\rm ^3He} 
                      - 2\frac{p_p}{p_n} d_2^p
\label{eq:he3ton}
\end{eqnarray}
where $p_n$ and $p_p$ correspond to the effective polarization of the neutron and proton inside $^3$He~\cite{Friar:1990vx}. 
This neutron extraction method is expected to be accurate at the 5\% level~\cite{CiofidegliAtti:1996cg}. The resonance and DIS contributions of $d_2^p$ 
were calculated using data from JLab experiment EG1b~\cite{Bosted:2006gp} for the proton spin structure function $g_1^p$ 
and the Hall B model~\cite{HallBmod} for $g_2^p$. A conservative uncertainty of 100\% on $g_2^p$ was taken into account in 
our systematics uncertainties.
\begin{figure}[h]
  \includegraphics[scale=0.55]{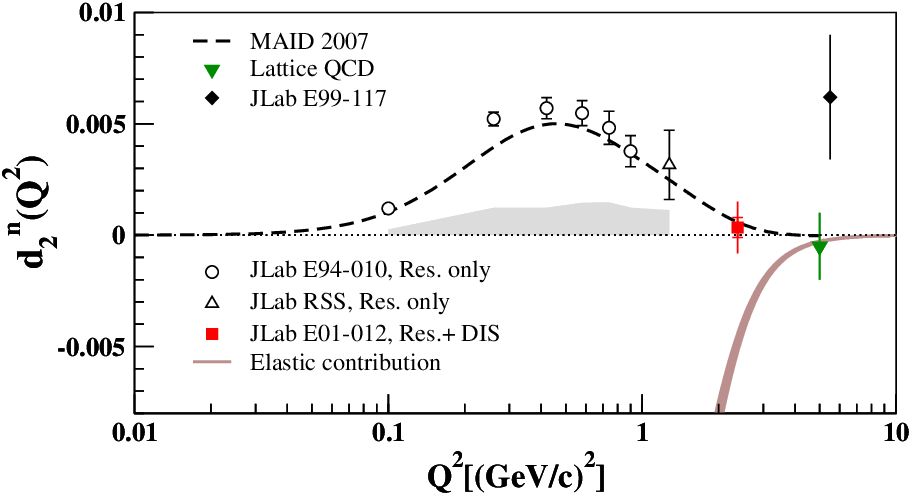}
  \caption{\label{fig_d2} (Color online) Result on {\it inelastic contribution} to the 
neutron $x^2$-weighted moment $d_2^n(Q^2)$ from E01-012. The  {\it elastic 
contribution} is displayed by the brown band. The inner (outer) error bar represents 
statistical (total) uncertainty. The {\it resonance contribution} to $d_2^n(Q^2)$ from 
JLab experiments E94-010~\cite{Amarian:2003jy} and RSS~\cite{Slifer:2008xu} are 
also shown: the error bars are statistical only and the grey band represents the 
experimental systematics uncertainties. To be compared with the  resonance contribution, 
we plotted the MAID model~\cite{Drechsel:2007if}. Also plotted are the {\it total} 
$d_2^n$ from SLAC E155x~\cite{Anthony:2002hy} and
JLab E99-117~\cite{Zheng:2003un,Zheng:2004ce} combined, and, the Lattice QCD 
prediction~\cite{Gockeler:2000ja}.}
\end{figure}

We extracted the {\it inelastic contribution} to $d_2^n$ at our four $Q^2$ values by adding the resonance and the DIS 
contributions (see Table~\ref{d2values}, where the results were multiplied by 10$^5$ for listing in the table). Including 
the $Q^2$-evolution from Ref.~\cite{Shuryak:1981pi}, we performed 
the weighted average and obtained $d_2^n = 0.00034 \pm 0.00045 \pm 0.00107$ for 
$\langle Q^2 \rangle= 2.4~\rm{(GeV/c)}^2$, as shown in Fig.~\ref{fig_d2}. The elastic contribution, shown separately, 
was evaluated using elastic form factors from Refs.~\cite{Arrington:2007ux,Bradford:2006yz} following the formalism of 
Ref.~\cite{Carlson:1998gf}. Uncertainties of 5\%, 1\%, 14\% and 2.5\% were assigned to the proton and neutron form factors 
$G_E^p$, $G_M^p$, $G_E^n$ and $G_M^n$, respectively. JLab experiments E94-010~\cite{Amarian:2003jy} and 
RSS~\cite{Slifer:2008xu} reported only the {\it resonance contribution} to $d_2^n$ and it can be seen that these data are 
in very good agreement with the MAID model~\cite{Drechsel:2007if}. Since $d_2(Q^2)$ is weighted by $x^2$, one would 
expect it to be dominated by the contribution coming from the resonance region, which sits at higher $x$ compared to the 
DIS region. Our data show the inelastic contribution to $d_2(Q^2)$ becoming very small by $Q^2=2~\rm{(GeV/c)}^2$,  as  also indicated by the MAID model. JLab E99-117~\cite{Zheng:2003un,Zheng:2004ce} evaluated $d_2(Q^2)$ at 
$\langle Q^2 \rangle = 5~\rm{(GeV/c)}^2$ including the previous data from SLAC experiment E155x~\cite{Anthony:2002hy}. 
The result shows $d_2(Q^2)$ large and positive, about 1.5$\sigma$ away from the Lattice QCD prediction~\cite{Gockeler:2005vw}. 
The trend of the 
experimental inelastic contributions at $Q^2 \le 2.4$ (GeV/c)$^2$  and the falloff of the elastic contribution appear to be in agreement with  the  Lattice QCD prediction at 5 GeV$^2$.

\begin{figure}[h!]
\includegraphics[scale=0.55]{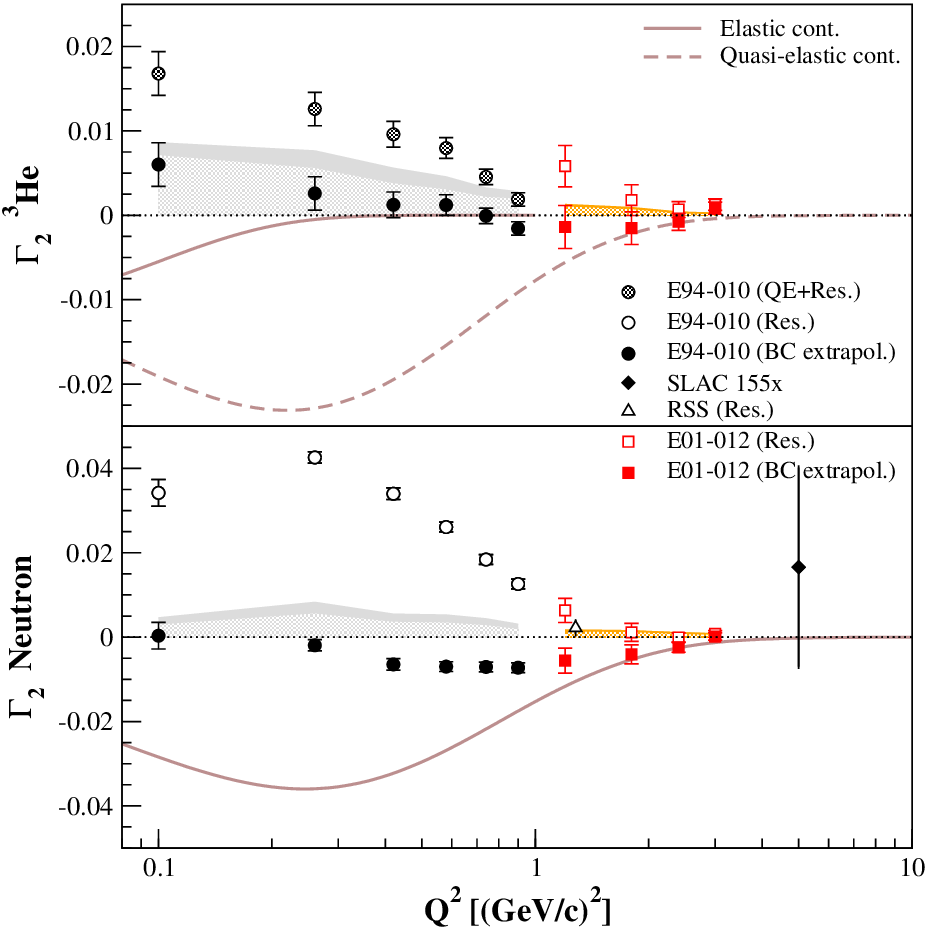}
\caption{\label{fig_bc} (Color online) The $^3$He (top panel) and neutron (bottom panel) 
$\Gamma_2$ integrals from JLab E01-012 (filled squares). The error bars are statistical only, 
the upper band represents the experimental systematics and the lower band the uncertainties 
on the unmeasured part of the BC sum. The open square data are the measured part of the 
integral as was performed by experiment E01-012. Also plotted are data from JLab experiments 
E94-010~\cite{Amarian:2003jy,Slifer:2008re} and RSS~\cite{Slifer:2008xu}, with also the measured part of the integral 
represented by open and shaded symbols and the extrapolated BC sum with filled symbols, and 
SLAC experiment E155x~\cite{Anthony:2002hy}. The elastic (solid line) and quasi-elastic 
(dashed line) contributions to the integrals are plotted.} 
\end{figure}

\section{The Burkhardt-Cottingham sum}
The Burkhardt-Cottingham (BC) sum rule~\cite{Burkhardt:1970ti} is a super-convergence  relation derived from a 
dispersion relation in which the virtual Compton helicity amplitude $S_2$ falls off to zero more rapidly than 
$\frac{1}{\nu}$ as $\nu \rightarrow \infty$. The sum rule is expressed as follows: 
\begin{eqnarray}
\Gamma_2(Q^2) \equiv \int_0^1 dx~g_2(x,Q^2) = 0,
\label{eq:bc}
\end{eqnarray}
and is predicted to be valid at all $Q^2$. It should be noted that the validity of the sum rule has been 
questioned~\cite{Jaffe:1989xx, Accardi:2009au}. Furthermore, the BC sum rule cannot be extracted from the OPE due to the non-existent 
$n=0$ expansion of $g_2$-moments. The  data for  $\Gamma_{2}(Q^2)$ at 5 (GeV/c)$^2$ from the SLAC E155x 
experiment showed that the BC sum rule is satisfied within a large uncertainty for deuteron. However, they found a 
violation of almost $3\sigma$ for the more precise proton measurement.

We separate the full  $\Gamma_{2}(Q^2)$ integral into DIS, resonance and elastic components  as follows:
\begin{eqnarray}
\Gamma_{2}(Q^2) & = & \Gamma_{2}^{DIS}(Q^2) + \Gamma_{2}^{Res}(Q^2) + \Gamma_{2}^{El}(Q^2) \nonumber \\
                       & =  & \int_0^{x_{min}} dx~g_2(x,Q^2) + 
                             \int_{x_{min}}^{x_{\pi}} dx~g_2(x,Q^2)  \nonumber \\
                       &    & +  \int_{x_{\pi}}^1 dx~g_2(x,Q^2). 
\label{eq:gamma2}
\end{eqnarray}
The variables $x_{min}$ and $x_{\pi}$ are the $x$ values corresponding to the invariant mass $W = 1.905$ GeV and to 
$W$ at pion threshold, respectively, at the given value of $Q^2$. We measured the $\Gamma_2^{Res}$ part in our 
experiment. The elastic contribution, $\Gamma_2^{El}$, was evaluated using the method as described in the previous section. 
The quasi-elastic contribution to the $^3$He BC sum was extracted from:
\begin{eqnarray}
 \Gamma_{2}^{{\rm ^3He},QE} = (p_n \Gamma_{2}^{n,EL} + 2 p_p \Gamma_{2}^{p,EL})/f
\label{eq:qegamma2}
\end{eqnarray}
where the $Q^2$-dependent scale factor $f = 1.12 + 0.65 Q^2$ was determined from comparison to the quasi-elastic 
data from E94-010. A relative uncertainty of 20\% was assumed for our evaluation of $\Gamma_{2}^{{\rm ^3He},QE}$ in 
order to include the total uncertainties of E94-010 data. Both the elastic and quasi-elastic contributions of the BC sum are 
shown in Fig.~\ref{fig_bc}.

There is not enough experimental data  currently available to evaluate $\Gamma_2^{DIS}$ in the $Q^2$ range covered by our 
experiment. Therefore, it is not possible to evaluate the full  $\Gamma_{2}(Q^2)$ integral to  test the BC sum rule 
without assumptions. Previously,  JLab Hall A experiment E94-010 evaluated the BC sum, using the $\Gamma^{WW}_2$ 
part for the  unmeasured DIS region, at six $Q^2$ values from 0.1 to 0.9 (GeV/c)$^2$. The same method was used here: 
$\Gamma_2^{WW}$ for $^3$He is calculated using our $g_1^{\rm ^3He}$ data~\cite{Solvignon:2008hk}. The extraction of the 
neutron $\Gamma_2$ integrals were done using the same method as described by Eq.~\ref{eq:he3ton}, using $g_1^p$ data 
from~\cite{Bosted:2006gp} and $g_2^p$ from Hall B model~\cite{HallBmod} to evaluate the proton $\Gamma_2^{WW}$ and 
$\Gamma_2^{Res}$ respectively. 
Figure~\ref{fig_bc}  shows $\Gamma_2^{Res}$ and the extrapolated BC sum for $^3$He and the neutron compared to 
the same quantities from the previous experiments E94-010~\cite{Amarian:2003jy, Slifer:2008re} and RSS~\cite{Slifer:2008xu}. 
It should be noted that RSS extracted their neutron result from the deuteron and the agreement with our data demonstrates that 
the nuclear corrections for deuteron and $^3$He are well understood. All results are in good agreement with the BC sum rule 
for $^3$He and within 2$\sigma$ from the neutron BC sum rule, as shown on the bottom panel of Fig.~\ref{fig_bc} and in 
Table~\ref{d2values} (the results were multiplied by $10^{5}$ for listing in the table). 

\section{Conclusion}
In summary, we have measured the inelastic contribution to the neutron $d_2(Q^2)$ matrix element at $<Q^2>=2.4$ (GeV/c)$^2$ 
and found it very small, in agreement with the Lattice QCD calculation. We also formed the $^3$He and neutron $\Gamma_2$ 
moments over the $Q^2$ range of 1.2 to 3.0 (GeV/c)$^2$. Our data show both moments to be small and to gradually decrease in 
magnitude with $Q^2$. The BC sum for $^3$He and the neutron was then evaluated from our data in the resonance region, adding 
the elastic contribution from elastic form factors and using $g_2^{WW}$ for the low $x$ unmeasured part of the integral. Our data 
confirmed the validity of the BC sum rule at the 1.5$\sigma$ level.

\begin{table*}[t]
\caption{\label{d2values} E01-012 results given at the scale of 10$^{-5}$.}
\vspace{0.2cm}
\centering
\begin{tabular}
{p{0.05\linewidth}p{0.07\linewidth}p{0.18\linewidth}p{0.18\linewidth}p{0.15\linewidth}p{0.18\linewidth}}
~ &  $Q^2$          & Resonance  &  DIS &   Elastic or QE  & Total \\
~ & (GeV/c)$^2$ &  (10$^{-5}$) &  (10$^{-5}$) &  (10$^{-5}$) &  (10$^{-5}$)\\
\hline\hline
\multirow{4}{*}{$d_2^n$} & 1.2 &   186 $\pm$  136  $\pm$ 156 & -2 $\pm$  6  $\pm$  3 & -2342 $\pm$  204 & -2158  $\pm$  136   $\pm$  257 \\
             & 1.8 &   -32 $\pm$   177  $\pm$ 107 &   1 $\pm$   9  $\pm$  5 & -1075 $\pm$    96 & -1105   $\pm$  177   $\pm$  144 \\
            & 2.4 &   -55 $\pm$   118  $\pm$ 101 &   3 $\pm$   7  $\pm$  4 & -468   $\pm$    40 & -520     $\pm$  118   $\pm$  109 \\
            & 3.0 &     80 $\pm$     88  $\pm$ 137 & 13 $\pm$   6  $\pm$  2 & -211   $\pm$    16 & -117     $\pm$  88     $\pm$  138 \\
\hline\hline
\multirow{4}{*}{$\Gamma_2^{\rm{^3He}}$} & 1.2  &  582   $\pm$   245  $\pm$    115 & -162  $\pm$    72  $\pm$    41   &  -558   $\pm$   31 &  -139  $\pm$    255  $\pm$   126 \\ 
                                        & 1.8  &  180   $\pm$   182  $\pm$      82 & -114  $\pm$    67  $\pm$    36   &  -219   $\pm$   12 &  -153  $\pm$    194  $\pm$     90 \\
                                        & 2.4  &    68   $\pm$     94  $\pm$      33 &   -55  $\pm$    38  $\pm$    18   &    -90   $\pm$     5 &    -77  $\pm$    101  $\pm$     37 \\
                                        & 3.0  &  127   $\pm$     68  $\pm$      23 &     -3  $\pm$    24  $\pm$      7   &    -40   $\pm$     2 &      84  $\pm$      72  $\pm$   24 \\  
\hline
\multirow{4}{*}{$\Gamma_2^n$} & 1.2  &   634  $\pm$   285   $\pm$   153   & -26  $\pm$    84  $\pm$   50  &  -1165   $\pm$   58  &  -558   $\pm$   297   $\pm$    171 \\ 
                       & 1.8  &   114  $\pm$   212   $\pm$   141   &   12  $\pm$    78  $\pm$   43  &    -532   $\pm$   27  &  -407   $\pm$   226   $\pm$    150 \\
                       & 2.4  &     -9  $\pm$   109   $\pm$     98   &   21  $\pm$    44  $\pm$   24  &    -253   $\pm$   13  &  -241   $\pm$   118   $\pm$    102 \\
                       & 3.0  &     78  $\pm$     79   $\pm$     76   &   65  $\pm$    28  $\pm$   10  &    -128   $\pm$     7  &     15    $\pm$     84   $\pm$      77 \\ 
\hline
\hline
\end{tabular}
\end{table*}

\begin{acknowledgments}
We would like to acknowledge the outstanding support from the Jefferson Lab Hall A 
technical staff. This work was supported by the NSF and DOE contract DE-AC05-06OR23177 
under which JSA, LLC operates JLab.
\end{acknowledgments}

%
\nocite{*}

\bibliography{e01012_g2}

\end{document}